\documentclass[times,10pt,twocolumn]{article} 
\usepackage{latex8}
\usepackage{times}

\usepackage{epsf}
\usepackage{amsmath}
\usepackage{latexsym}
\usepackage{amssymb}
\usepackage{rotating}
\usepackage{setspace}
\usepackage{graphicx}
\usepackage{latexsym}
\usepackage{amsfonts}
\usepackage{subfigure}
   
\newtheorem{theorem}{Theorem}[section]
\newtheorem{lemma}[theorem]{Lemma}

\newcommand{\N}{\mathbb N}
\newcounter{linecounter}
\newcommand{\linenumbering}{(\arabic{linecounter})}
\renewcommand{\line}[1]{\refstepcounter{linecounter}
\label{#1}
\linenumbering}
\newcommand{\resetline}{\setcounter{linecounter}{0}}

\newif\ifcode
\codefalse

\codetrue
\usepackage{macros}
\ifcode
\usepackage[ruled]{algorithm}
\usepackage{algpseudocode}
\usepackage{ioa_code}
\fi

\newcommand{\remove}[1]{}

\begin{document}

\title{Distributed Slicing in Dynamic Systems}

\author{Antonio Fern\'andez\footnotemark[1] ~
Vincent Gramoli\footnotemark[2] ~
Ernesto Jim\'enez\footnotemark[3] ~
Anne-Marie Kermarrec\footnotemark[2] ~
Michel Raynal\footnotemark[2]\\
\and
\footnotemark[1]{}~~Universidad Rey\\ Juan Carlos,\\ 28933 M\'ostoles, Spain\\ anto@gsyc.escet.urjc.es
\and
\footnotemark[2]{}~~IRISA, INRIA\\ Universit\'e Rennes 1,\\ 35042 Rennes, France\\ \{vgramoli,akermarr,raynal\}@irisa.fr
\and
\footnotemark[3]{}~~Universidad Polit\'ecnica\\ de Madrid,\\ 28031 Madrid, Spain\\ ernes@eui.upm.es
}

\maketitle
\thispagestyle{empty}

\begin{abstract}
 	Peer to peer (P2P) systems are moving from application specific architectures to 
	a generic service oriented design philosophy.
	This raises interesting problems in connection with providing useful
	P2P middleware services 
	capable of dealing with resource assignment
	and management in a large-scale, heterogeneous and unreliable environment.
	The slicing service, has been proposed to allow for an
	automatic partitioning of P2P networks into groups (slices) that
	represent a controllable amount of some resource and that are also relatively
	homogeneous with respect to that resource.
	In this paper we propose two gossip-based algorithms to solve the distributed slicing problem.
	The first algorithm speeds up an existing algorithm 
	sorting a set of uniform random numbers.
	The second algorithm 
	statistically  approximates
	the rank of nodes in the ordering.
	The scalability, efficiency and resilience to dynamics of both algorithms 
	rely on their gossip-based models.
	These algorithms are proved viable theoretically and experimentally.
\end{abstract}


\Section{Introduction}

\SubSection{Context and Motivations}

The peer to peer (P2P) communication paradigm has now become the prevalent
model to build large-scale distributed applications.
%
%
On the one hand, P2P protocols 
integrate into
platforms on top of which several applications, with various requirements,
may cohabit. 
This leads to the interesting issue of resource 
assignment or how to allocate a set of nodes for a given application. 
Examples of applications for such a service are telecommunication,
testbed platform~\cite{BBC+04}, 
or desktop-grid-like applications~\cite{A04}.
%
On the other hand, P2P systems should be able to balance the 
load taking into account that capabilities are heterogeneous 
at the peers~\cite{SGG02,BSV03,SR06}.
%
%
This heterogeneity has some drawbacks.
Completely decentralized P2P
application, like the original Gnutella~\cite{gnutella}, suffered from congestion when 
applied to large-scale systems because nodes with a low bandwidth 
capability were queried.
Consequently, file sharing applications~\cite{kazaa,gpd} 
tend
to request ultrapeers/supernodes (peers with larger lifetime and 
bandwidth capabilities), more often than regular peers.  
P2P protocols must identify efficiently and accurately peers
with specific capabilities.

Large scale dynamic distributed systems consist of many 
participants that can join and leave 
at will. Identifying peers in such systems that have a similar level
 of power or capability (for instance, in terms of bandwidth, 
processing power, storage space, or uptime) in a completely 
decentralized manner is a difficult task. It is even harder 
to maintain this information in the presence of churn. 
 Due to the intrinsic dynamics of contemporary P2P
 systems it is impossible to obtain accurate information about 
the capabilities (or even the identity) of the system participants. 
Consequently, no node is able to maintain accurate information about 
all the nodes. This  disqualifies  centralized approaches.  

The slicing service~\cite{JK06} enables peers 
to self-organize into a partitioning,
where partitions (slices) are connected overlay networks that represent
a given percentage of some resource.
The slicing is ordered in the sense that peers are sorted
according to their capabilities expressed by an attribute value.
Building upon the work on ordered slicing
of~\cite{JK06}, here we focus on the issue of 
\textit{accurate} slicing. 
That is, we focus on improving quality by slicing the network accurately,
and improving stability of slices by minimizing the impact of the churn.
%
The distributed slicing problem we tackle in this paper consists in 
ranking nodes depending on their relative capability, slicing the network 
depending on these capabilities and, most importantly, readapting
the slices continuously to cope with system dynamism.  

\SubSection{Contributions}

The paper presents two distributed algorithms to slice the nodes
according to their capability, reflected by an attribute value. 
Theses algorithms are robust and lightweight due to their
gossip-based communication pattern.
%
%
The first algorithm of the paper builds upon the ordered 
slicing algorithm proposed in \cite{JK06} that we call the JK algorithm in the
sequel of this paper.
This algorithm speeds up the convergence of JK by locally computing a 
disorder measure so that a peer chooses the neighbor to communicate 
with in order to  maximize the chance of decreasing the global disorder 
measure.
%

Then, we identify two issues that prevent accurate slicing and motivate 
us to find an alternative approach to this algorithm and JK.
First, the slicing might be inaccurate.
Random values are used
 to calculate which slice a node belongs to. 
The accuracy of the slicing 
 fully depends on the uniformity of the random value
 spread between 0 and 1. 
(e.g., the proportion of 
random values between 0.8 and 1 should be ideally 
20\% of the nodes). 
%
%
Second, the previous algorithms suffer from churn an dynamism
when correlated with the attribute values. For example, if the peers are 
sorted according to their connectivity potential, 
a portion of the attribute space (and therefore the random value space) might be 
suddenly affected. 
The consequence is to skew the distribution of random values towards high or low 
values.

The second algorithm is an alternative algorithm solving these two issues  
by approximating locally the rank of the nodes, without
using random values.
  The basic idea is that each node periodically estimates its rank 
along the attribute axis depending on the attributes it have seen so far.
Based on continuously aggregated information, 
the node can determine the
slice it belongs to with a decreasing error margin.  
We show that this algorithm provides accurate estimation and recovery ability in presence 
of attributes-correlated churn at the price of a slower convergence.


\SubSection{Outline}
The rest of the paper is organized as follows: Section \ref{related} surveys some related work. 
The system model is presented in Section \ref{model}. 
The first contribution of an improved ordered slicing algorithm based 
on random values is presented  in Section~\ref{JK+} and the second algorithm 
based on dynamic ranking in Section~\ref{sec:ranking}. 
Section~\ref{conclusion} concludes the paper.

\Section{Related Work}
\label{related}
Most of the solutions proposed so far for ordering nodes come
from the context of databases~\cite{DNS91,IRV89},
where parallelizing query executions is used to improve efficiency. 
A large majority of the solutions in this area rely on centralized gathering
or all-to-all exchange, which makes them unsuitable 
for large-scale networks.

Other related problems are the selection problem and the $\phi$-quantile search.
The selection problem~\cite{FR75} aims at determining the $i^{th}$
smallest element with as few comparisons as possible.
The $\phi$-\emph{quantile} search (with 
$\phi \in (0,1]$) is the problem to find among $n$ elements the $(\phi n)^{th}$ element.
Even though these problems look similar to our problem, 
they aim 
at finding a specific node among all, while the distributed slicing problem aims
at solving a global problem where each node maintains a piece of
information.  
Additionally, solutions to the quantile search problem like the one presented 
in~\cite{KDG03} use an approximation of the system size. The same holds 
for the algorithm in~\cite{SDCM06}, which uses similar ideas to determine the distribution
of a utility in order to isolate peers with high capability---i.e., super-peers.

As far as we know, the distributed slicing problem was studied in a P2P system 
for the first time in~\cite{JK06}. 
In this paper, every node draws independently and uniformly a random value in the
interval $(0,1]$. Each of these values serve as an estimate of normalized index $k/n$
for the node with the $k^{th}$ smallest attribute value.

\Section{Model and Problem Statement}
\label{model}

\SubSection{System model}

We  consider a  system $\Sigma$ containing  a set
of $n$ uniquely identified nodes.
(The value $n$ may vary over time.)
The set of identifiers is denoted by $I \subset \N$.
Each node can leave and new nodes can join the system at any  
time, thus the number of nodes is a function of time.
Nodes may also crash. In this paper,
we do not differentiate between a crash and a voluntary node departure. 

Each node $i$ maintains a fixed attribute value $a_i \in \N$, reflecting the node
capability according to a specific metric.
These attribute values over the network might have an arbitrary
skewed distribution.
Initially, a node
has no global information neither about the structure or size of the system nor
about 
the attribute values of the other nodes. 

We can define a total ordering over the nodes based on their attribute value,
with the node identifier used to break ties.
Formally, we let $i$ precede $j$ if and only if $a_i < a_j$, or
$a_i = a_j$ and $i < j$. We refer to this totally ordered sequence as the
\emph{attribute-based sequence}, denoted by $A.\ms{sequence}$. The attribute-based
rank of a node $i$, denoted by $\alpha_i \in \{1,...,n\}$, is defined as the index
of $a_i$ in $A.\ms{sequence}$. 
\remove{
For instance, let us consider three nodes: 1, 2, and 3, with three
different attribute values $a_1 = 50$, $a_2 = 120$, and $a_3 = 25$.
In this case, the attribute-based rank of node $1$ would be
$\alpha_1 = 2$.
In the rest of the paper, we assume that nodes are sorted according to a single 
attribute and that each node belongs to a unique slice.
The sorting along several attributes is out of the scope of this paper.
} 

\SubSection{Distributed Slicing Problem}\label{ssec:pb}

Let ${\cal S}_{l,u}$ denote the \emph{slice} containing every node $i$ whose
normalized rank, namely $\frac{\alpha_{i}}{n}$, satisfies $l < \frac{\alpha_{i}}{n}
\leq u$
where $l\in [0,1)$ is the slice lower boundary and $u\in (0,1]$ is 
the slice upper boundary so that all slices represent adjacent intervals $(l_1, u_1], (l_2, u_2]$...
Let us assume that we partition the interval $(0,1]$ using a set of slices,
and this partitioning is known by all nodes.
The distributed slicing problem requires each node 
to determine the slice it currently belongs to.
Note that the problem stated this way is similar to the
ordering problem, where each node has to determine its own index in
$A.\ms{sequence}$.
However, the reference to slices introduces special requirements related to
stability and fault tolerance, besides, it allows for future generalizations
when one considers different types of categorizations.

\begin{figure}[t]
\centering\includegraphics[scale=0.6]{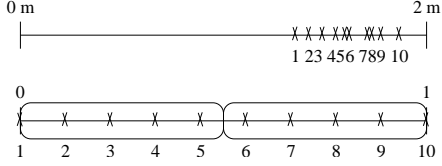}
\caption{Slicing of a population based on a height attribute.}
\label{fig:size}
\end{figure}

 Figure~\ref{fig:size} illustrates an example 
of  a population of 10 persons, to be sorted 
against their height.
A partition of this population could be defined by two slices of the same
size: the group of short persons, and the 
group of tall persons. This is clearly an example 
where the distribution of attribute values is skewed towards 2 meters.
 The rank of each person in
the population and the two slices are represented on the bottom axis. Each
person is represented as a small cross on these axes.\footnote{Note that the shortest (resp. largest) rank is represented by a cross at the 
extreme left (resp. right) of the bottom axis.} 
Each slice is represented as an oval.  The slice $S_1 = {\cal
S}_{0,\frac{1}{2}}$ contains the five shortest persons and 
the slice $S_2 = {\cal S}_{\frac{1}{2},1}$ contains the five tallest
persons.

\remove{
Observe that another way of partitioning the population could be to define the
group of short persons as that containing all the persons shorter than a 
predefined measure (e.g., $1.65m$) and the group of tall persons as that containing
the persons taller than this measure.
However, this way of partitioning would most certainly lead 
to have empty groups that contains no nodes (while a
slice is almost surely non-empty). 
} 

Since the distribution of attribute values is
unknown and hard to predict, defining relevant groups is a difficult task.  For
example, if the distribution of the human heights were unknown, then the persons
taller than $1m$ could be considered as tall and the persons shorter than $1m$
could be considered as short. In this case, the first of the two groups would be 
empty, while the second of the two groups would be as big as the whole system.
Conversely, slices partition the population  
into subsets representing a predefined portion of this population.
Therefore, in the rest of the paper, we consider slices as defined as a proportion of the
network.

\SubSection{Facing Churn}
Node churn, that is, the continuous arrival and departure of nodes is an intrinsic characteristic 
of P2P systems and may significantly impact  the outcome, and more specifically the accuracy 
of the slicing algorithm.
The easier case is when  the distribution of the attribute values of the departing and
arriving nodes are identical.
In this case, in principle, the arriving nodes must find their slices, but
the nodes that stay in the system  are mostly able to keep their slice assignment.
Even in this case however, nodes that are close to the border of a slice
may expect frequent changes in their slice due to the variance of the
attribute values, which is non-zero for any non-constant distribution.
If the arriving and departing nodes have different attribute distributions,
 so that the distribution in the actual network of live nodes keeps changing,
then this effect is amplified. However, we believe that this is a realistic assumption
to consider that the churn may be correlated with some specific values (for example 
if the considered attribute is uptime or connectivity).

\Section{Dynamic Ordering by Exchange of Random Values}\label{sec:ordering}
\label{JK+}

This section proposes an  algorithm for the distributed slicing 
problem improving upon the original JK algorithm \cite{JK06},
 by considering a local measure of the global disorder function.
 In this section we present the algorithm along with the corresponding analysis and 
simulation results.

\SubSection{On Using Random Numbers to Sort Nodes}

This Section presents the algorithm built upon JK. 
We refer to this algorithm as \emph{mod-JK} (standing for modified JK).
In JK,
each node $i$
 generates a real number $r_i\in (0,1]$ independently and uniformly at random.
The key idea is to sort these random numbers with 
respect to the attribute values by swapping (i.e., exchanging) these random numbers between nodes,
so that if $a_{i} < a_{j}$ then $r_{i} < r_{j}$. 
Eventually,  the attribute values (that are fixed) and the 
random values (that are exchanged)  should be sorted in the same order.
That is, each node would like to obtain the $x^{th}$ largest random number if 
it owns the $x^{th}$ largest attribute value.
Let $R.\ms{sequence}$ denote the \emph{random sequence} obtained 
by ordering all nodes according to their random number.
Let $\rho_i(t)$ denote the index of node $i$ in $R.\ms{sequence}$ at time
$t$. When not required, the time parameter is omitted.

\remove{
To illustrate the above ideas, consider that nodes 1, 2, and 3 
from the previous example
have three
distinct random values: $r_1 = 0.85$, $r_2 = 0.1$, and $r_3 = 0.35$.
In this case, the index $\rho_1$ of node $1$ would be $3$. Since the
attribute values are $a_{1}=50$, $a_{2}=120$, and $a_{3}=25$, the algorithm
must achieve the following final assignment of random numbers: $r_{1}=0.35$,
$r_{2}=0.85$, and $r_{3}=0.1$.
} 

Once sorted, the random values are used to determine the portion of the network a peer belongs to.

\SubSection{Definitions}

Every node $i$ keeps track of some neighbors and their age. 
The \emph{age} of neighbor $j$ is a timestamp, $t_j$, set to 0 when $j$ becomes a 
neighbor of $i$.
Thus, node $i$ maintains an array
containing the id, the age,
the attribute value, and the 
random value of its neighbors.  This array, denoted ${\mathcal N}_i$, is 
called the \emph{view} of node $i$.
The views of all nodes have the same size, denoted by $c$.  
A node $i$ participates in the algorithm by exchanging its rank with a \emph{misplaced} 
neighbor in its view.  Neighbor $j$ is misplaced 
if and only if
$(a_j - a_i)(r_j - r_i)<0$.
In~\cite{JK06}, a measure of the relative
disorder of sequence $R.\ms{sequence}$ with respect to sequence
$A.\ms{sequence}$ was introduced. We call it the 
\emph{global disorder measure (GDM)} and it is defined, for any time $t$, as
$\ms{GDM}(t) = \frac{1}{n}\sum_{i}(\alpha_i - \rho(t)_i)^2.$
The minimal value of GDM is 0, which is obtained
when $\rho(t)_i = \alpha_i$ for all nodes $i$.  In this case the 
attribute-based index of a node is equal to its random value index, indicating
that random values are ordered.

\SubSection{Improved Ordering Algorithm}

In this algorithm, each node $i$ searches its own view ${\mathcal N}_i$
 for misplaced neighbors. Then, 
one of them is chosen to swap  random value with. This process is repeated
until there is no global disorder.
In this version of the algorithm, we provide each node with  the capability
of measuring disorder locally. This leads to  a new
heuristic for each node to determine 
the neighbor to exchange with which decreases most the disorder.
Referring to this disorder measure as a criterion,
the decrease of the global criterion is related to the decrease of local 
criteria, similarly to~\cite{ADGR05}.
%

For a node $i$ to evaluate the gain of exchanging with a node $j$
of its current view ${\mathcal N}_i$,
we define its \emph{local disorder measure} (abbreviated \emph{LDM}$_i$).  Let $LA.\ms{sequence}_i$ and 
$LR.\ms{sequence}_i$ be the local attribute sequence and the local 
random sequence of node $i$, respectively.   These sequences are computed locally by $i$ 
using the information ${\mathcal N}_i \cup \{i\}$.  Similarly to $A.\ms{sequence}$
and $R.\ms{sequence}$, these are the sequences of neighbors where each node
is ordered according to its attribute value and random number, respectively.
Let, for any $j\in {\cal N}_i \cup \{i\}$, $\ell\rho_j(t)$ and $\ell\alpha_j(t)$ be the 
indices of $r_j$ and $a_j$ in sequences $LR.\ms{sequence}_i$ and
$LA.\ms{sequence}_i$, respectively, at time $(t)$.
At any time $t$, the local disorder measure of node $i$ is defined as:
\begin{eqnarray}
\ms{LDM}_i(t) &=& \frac{1}{c+1}\sum_{j\in {\mathcal N}_i(t)\cup \{i\}}(\ell\alpha_j(t) - \ell\rho_j(t))^2.\label{eq:ldm}
\end{eqnarray}
We denote by $G_{i,j}(t+1) = \ms{LDM}_i(t) - \ms{LDM}_i(t+1)$, the reduction 
on this measure that $i$ obtains after exchanging its random value with node 
$j$ between time $t$ and $t+1$.

The heuristic used chooses for node $i$ the misplaced neighbor $j$ that
maximizes $G_{i,j}(t+1)$.

\paragraph{Sampling uniformly at random.}

The algorithm relies on the fact that potential misplaced nodes are found
so that they can swap their
random numbers thereby increasing order. 
If the global disorder is high,  it is very likely that
any given node has misplaced neighbors in its view to exchange with.
%
Nevertheless, as the system gets ordered, it becomes more unlikely for a node
$i$ to have misplaced neighbors.
In this stage the way the view is composed plays a crucial role: if fresh
samples from the network are not available, convergence can be slower than
optimal.

Several protocols may be used to provide a random and dynamic sampling 
in a P2P system such as Newscast~\cite{JMB05}, Cyclon~\cite{VGS05} or Lpbcast~\cite{JGKS04}.
They differ mainly by their \textit{closeness}
to the uniform random sampling of the neighbors and the way they handle churn.
In this paper,
we chose to use a variant of the Cyclon
protocol, to construct and update the views, as it is reportedly
the best approach to achieve a uniform random neighbor set for all nodes~\cite{I05}.
%


\paragraph{Description of the algorithm.}
\label{sec:modifiedjk}

\sloppy{The algorithm is presented in Figure~\ref{alg:rand}.  The active thread at node $i$
runs the membership (gossiping) procedure ($\lit{recompute-view}()_i$)
and the 
exchange of random values periodically.
As motivated above, the membership procedure is similar to 
the Cyclon algorithm.}
This variant of Cyclon
exchanges all entries of the view at each step and uses two additional parameters: 
the attribute value and the random value. 
For the detailed pseudocode, please refer to the full version of 
this paper~\cite{FGJKR06b}.

\begin{figure}[t]
\centering{
\fbox{
\begin{minipage}[ht!]{150mm}
\footnotesize
\renewcommand{\baselinestretch}{1.5}
\resetline
\begin{tabbing}
aaaaA\=aaaaaA\=aaaaaaA\kill
{\bf Initial state of node $i$} \\
\line{L00} \> $\ms{period}_{i}$, initially set to a constant; \\
$r_i$, a random value chosen in $(0,1]$; 
$a_i$, the attribute value;   \\
$\ms{slice}_i \gets \bot$, the slice $i$ belongs to;
${\mathcal N_i}$, the view; \\
$\ms{gain}_{j'}$, a real value indicating the gain achieved by \\
\T exchanging with $j'$; \\
$\ms{gain-max} = 0$, a real.
\\ ~ \\

{\bf Active thread at node $i$} \\
\line{M01} \> $\act{wait}(\ms{period_i})$ \\
\line{M02} \> $\act{recompute-view}()_i$ \\
\line{M03} \> {\bf for} $j' \in {\mathcal N}_i$ \\
\line{M04} \> \T {\bf if} $\ms{gain}_{j'} \geq \ms{gain-max}$ {\bf then} \\
\line{M05} \> \T \T $\ms{gain-max} \gets \ms{gain}_{j'}$ \\
\line{M06} \> \T \T $j \gets j'$ \\
\line{M07} \> {\bf end for} \\
\line{M08} \> $\act{send}(\lit{REQ}, r_i, a_i)$ to $j$ \\
\line{M09} \> $\act{recv}(\lit{ACK}, r_j')$ from $j$ \\
\line{M10} \> $r_j \gets r_j'$ \\
\line{M11} \> {\bf if} $(a_j - a_i)(r_j - r_i) < 0$ {\bf then} \\
\line{M12} \> \T $r_i \gets r_j$ \\
\line{M13} \> \T $\ms{slice}_i \gets {\cal S}_{l,u}$ such that $l < r_i \leq u$ \\
 ~ \\
 
 {\bf Passive thread at node $i$ activated upon reception} \\
\line{R01} \> $\act{recv}(\lit{REQ}, r_j, a_j)$ from $j$ \\
\line{R02} \> $\act{send}(\lit{ACK}, r_i)$ to $j$ \\
\line{R03} \> {\bf if} $(a_j - a_i)(r_j - r_i) < 0$ {\bf then} \\
\line{R04} \> \T $r_i \gets r_j$ \\
\line{R05} \> \T $\ms{slice}_i \gets {\cal S}_{l,u}$ such that $l < r_i \leq u$
 
\end{tabbing}
\normalsize
\end{minipage} 
}
\caption{Dynamic ordering algorithm.}
\label{alg:rand}
}
\end{figure}

The algorithm for exchanging random values from node $i$ starts by measuring 
the ordering that can be gained by swapping with each neighbor (Lines~\ref{M03}--\ref{M07}).
Then, $i$ chooses the neighbor $j \in {\mathcal N}_i$ that maximizes gain 
$G_{i,k}$ for any of its neighbor $k$. Formally, $i$ finds 
$j\in {\mathcal N}_i$ such that for any $k\in{\mathcal N}_i$, we have
$$G_{i,j}(t+1) \geq G_{i,k}(t+1).$$  
\noindent In Figure~\ref{alg:rand} of node $i$, we refer to $\ms{gain}_j$ as the value of
$\ell\alpha_i(t) \ell\rho_j(t) + \ell\alpha_j(t) \ell\rho_i(t) - \ell\alpha_j(t)
\ell\rho_j(t)$. This expression is directly obtained from equation~(\ref{eq:ldm}), see
the full version of this paper~\cite{FGJKR06b} for furthter details.

From this point on, $i$ exchanges its random value $r_i$ with the random value $r_j$ of 
node $j$ (Line~\ref{M10}). The passive threads are executed upon reception of a message.
In Figure~\ref{alg:rand}, when $j$ receives the random value $r_i$ of node 
$i$, it sends back its own random value $r_j$ for the exchange to occur (Lines~\ref{R01}--\ref{R02}).
Observe that the attribute value of $i$ is also sent to $j$, so that $j$ can
check if it is correct to exchange before updating its own random number (Lines~\ref{R03}--\ref{R04}). Node
$i$ does not need to receive attribute value $a_{j}$ of $j$, since $i$ already has this 
information in its view and the attribute value of a node never changes over time.

\SubSection{Analysis of Slice Inaccuracy}
\label{sec:anal}

In mod-JK, as in JK, the current random number
$r_{i}$ of a node $i$ determines the slice $s_{i}$ of the node. The
objective of both algorithms is to reduce the global disorder as quickly as
possible.  Algorithm mod-JK consists in choosing one neighbor among the possible neighbors
that would have been chosen in JK, plus the GDM of JK has been shown to fit an exponential 
decrease. Consequently mod-JK experiences also an exponential decrease of the global disorder.
Eventually, JK and mod-JK 
ensure that the disorder has fully disappeared. For further information, please refer to~\cite{JK06}.

However, the accuracy of the slices heavily depends on the uniformity of the random value
spread between 0 and 1.  It may happen, that the distribution of the random values
is such that some peers decide upon a wrong slice. Even more problematic is the fact that this situation
is unrecoverable unless a new random value is drawn for all nodes. 
This may be considered as an inherent limitation of
the approach.
For example, consider a system of size 2, where nodes 1 and 2 have the
random values $r_1=0.1$, $r_2=0.4$. If we are interested in creating
two slices $S_1$ and $S_2$ of equal size ($S_1 = {\cal
S}_{0,\frac{1}{2}}$ and $S_2 = {\cal
S}_{\frac{1}{2},1}$), 
both nodes will wrongly believe to belong to the same slice $S_1$, since $r_1$ and $r_2$ belong to $(0,\frac{1}{2}]$.
This wrong estimate holds even after perfect ordering of the random values.

Therefore, an important step is to characterize the inaccuracy of slice
assignment and how likely it may happen. 
To this end, we bound the deviation 
of random values distribution from the mean, and we lower bound the 
probability that this happen with only two slices.
The following result bounds, with high 
probability, the number of nodes that can be misplaced in the system.
For the proof of Lemma~\ref{lem:toprove} please refer to the full version of this paper~\cite{FGJKR06b}.

\begin{lemma}\label{lem:toprove}
For any $\beta \in (0,1]$, a slice $S_p$ of length  $p \in (0,1]$ has a
number of peers 
$X \in [(1-\beta)np, (1+\beta)np]$ with probability at least 
$1-\epsilon$ as long as $p  \geq \frac{3}{\beta^2 n} \ln (2/\epsilon)$.
\end{lemma}

To measure the effect discussed above during the simulation experiments, 
we introduce the slice 
disorder measure (SDM) as the sum over all nodes $i$ of the
distance between the slice $i$ actually belongs to and the slice $i$ believes
it belongs to.
For example (in the case where all slices have the same size), 
if node $i$ belongs to the $1^{st}$ slice (according to its attribute 
value) while it thinks it belongs to the $3^{rd}$ slice (according to its rank 
estimate) then the distance for node $i$ is $|1-3| = 2$.
Formally, for any node $i$, let $S_{u_i, l_i}$ be the actual correct slice 
of node $i$ and let $S_{\hat{u_i}, \hat{l_i}}(t)$ be the slice $i$ estimates
as its slice at time $t$.  The slice disorder measure is defined as:
$$\ms{SDM}(t) = \sum_{i}\frac{1}{u_i - l_i}\left|\frac{u_i+l_i}{2}-\frac{\hat{u_i}+\hat{l_i}}{2}\right|.$$
$\ms{SDM}(t)$ is minimal (equals 0) if for all nodes $i$, 
we have $S_{\hat{u_i}, \hat{l_i}}(t) = S_{u_i, l_i}$.

\SubSection{Simulation Results}\label{sec:simu2}

We present simulation results using PeerSim~\cite{JMB04}, using a simplified
cycle-based simulation model, where all messages exchanges are atomic,
so messages never overlap.
First, we compare the performance of the two algorithms: JK and mod-JK.
Second, we study the impact of concurrency that is ignored by the cycle-based
simulations.

\paragraph{Performance comparison.}

We compare the time taken by these algorithms 
to sort the random values according to the attribute values (i.e., the node with 
the $j^{th}$ largest attribute value of the system value obtains the $j^{th}$ random value).
In order to evaluate the convergence speed of each algorithm, we use the slice
disorder measure as defined in Section~\ref{sec:anal}.

\begin{figure}[t]
  \begin{center}
    \label{fig:convergence1}
          \resizebox{2.8in}{!}{\includegraphics[scale=0.5,angle=270,clip=true]{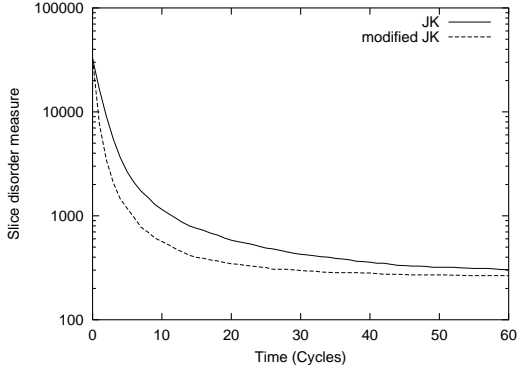}}
    \caption{Slice disorder measure over time.}
  \end{center}
\end{figure}

We simulated $10^4$ participants in 100 equally sized slices (when unspecified), each with 
a view size $c=20$.  
Figure~\ref{fig:convergence1} presents the evolution of
the slice disorder measure over time for JK, and mod-JK.


Figure~\ref{fig:convergence1} shows the slice disorder measure to compare
the convergence speed of our algorithm to that of JK with 10 equally sized slices.
Our algorithm converges significantly faster than JK.
Note that none of the algorithm reaches zero SDM, since they are both based on
the same idea of sorting randomly generated values.
Besides, since they both used an identical set of randomly generated values,
both converge to the same SDM. Note that for the sake of fairness, JK and mod-JK 
are compared using the same underlying view management protocol in our 
simulation: the variant of Cyclon.

\paragraph{Concurrency.}

The simulations are cycle-based and at each cycle an 
algorithm step is done atomically so that no other execution is concurrent.
More precisely, the algorithms are simulated such that in each cycle, each node updates its view 
before sending its random value or its attribute value.  Given this implementation, the 
cycle-based simulator does not allow us to realistically simulate concurrency, and a
drawback is that view is up-to-date when a message is sent.  In the following 
we artificially introduce concurrency (so that view might be out-of-date) into the 
simulator and show that it has only a slight impact on the convergence speed.

Adding concurrency raises some realistic problems due to
the use of non-atomic push-pull~\cite{JGKS04} in each message exchange.
That is, concurrency might lead to other problems because of
the potential staleness of views: unsuccessful swaps due to useless messages.  
Technically, the view of node $i$ might 
indicate that $j$ has a random value $r$ while this value is no longer 
up-to-date.  This happens if $i$ has lastly updated its view before $j$ swapped
its random value with another $j'$.
Moreover, due to asynchrony, it could happen that by the time a message is received
this message has become useless.
Assume that node $i$ sends its random value $r_i$ to $j$
in order to obtain $r_j$ at time $t$ and $j$ receives it by time $t+\delta$. 
With no loss of generality assume $r_i > r_j$.  Then if $j$ swaps its random value with
$j'$ such that $r_j'>r_i$ between time $t$ and $t+\delta$, then the message of $i$ 
becomes \emph{useless} and the expected swap does not occur (we call this an \emph{unsuccessful swap}).

\begin{figure*}
  \begin{center}
    \subfigure[]
    { 
      \label{fig:swap}
      \resizebox{2.6in}{!}{\includegraphics[scale=0.5,angle=0,clip=false,bb=40 72 260 150,viewport=0 0 260 150
      ]{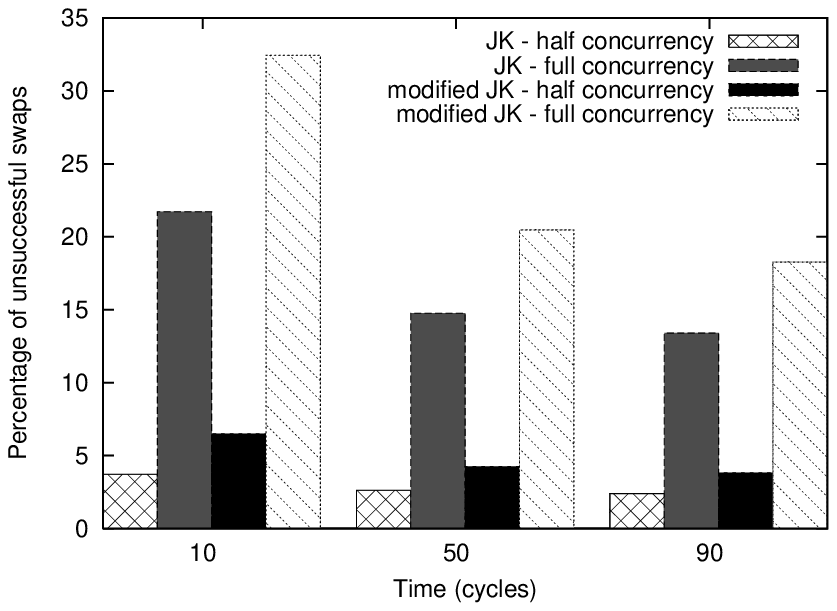}}
    } 
    \hspace{1cm}
    \subfigure[]
    { 
      \label{fig:sdm_count}
      \resizebox{2.6in}{!}{\includegraphics[scale=0.02,angle=270,clip=false,bb=510 50 0 550]{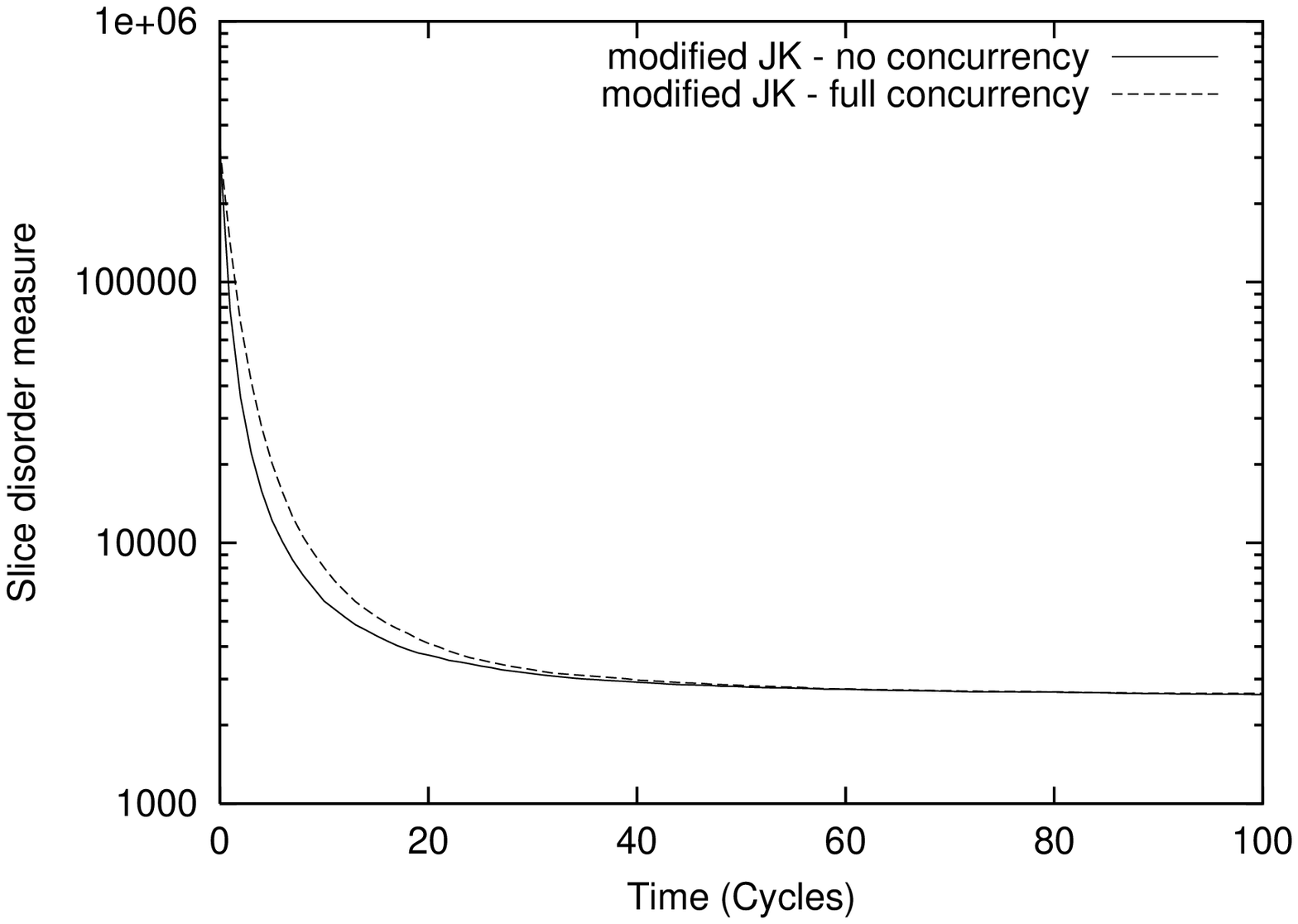}}
    }    
    \caption{
       (a) Percentage of unsuccessful swaps.
       (b) Convergence speed under high concurrency.}
  \end{center}
\end{figure*}
%

Figure~\ref{fig:sdm_count} indicates the impact of concurrent message exchange
on the convergence speed.
while
Figure~\ref{fig:swap} shows the amount of useless messages that are sent.
Now, we explain how the concurrency is simulated.
Let the \emph{overlapping messages} be a set of messages that mutually overlap: 
it exists, for any couple of overlapping messages, at least one instant at 
which they are both in-transit.  
For each algorithm we simulated \textit{(i)}~full concurrency: in a given 
cycle, all messages are overlapping messages; and \textit{(ii)}~half 
concurrency: in a given cycle, each message is an overlapping message with 
probability $\frac{1}{2}$.
Generally, we see that increasing the concurrency increases the number of
useless messages. Moreover, in the modified version of JK, more messages are 
ignored than in the original JK algorithm.  This
is due to the fact that some nodes (the most misplaced ones) are more likely 
targeted which increases the number of concurrent messages arriving at the 
same nodes.  Since a node $i$ ignored more likely a message when it receives
more messages during the same cycle, it comes out that concentrating
message sending at some targets increases the number of useless messages.
%


Figure~\ref{fig:sdm_count} compares the convergence speed under full concurrency 
and no concurrency.  
Full-concurrency 
impacts on the convergence speed very slightly.

\Section{Dynamic Ranking by Sampling of Attributes}\label{sec:ranking}
\label{dynamicranking}
In this section we propose an alternative algorithm for the distributed slicing
problem. This algorithm circumvents 
the problems identified in
the previous approach by continuously ranking nodes based on observing attribute value
information. 
Besides, this algorithm
is not sensitive to churn even if it is correlated with attribute values. 
In the remaining part of the paper we refer to this new algorithm 
as the ranking algorithm while referring to JK and mod-JK as the ordering algorithms.

\paragraph{Impact of dynamics correlated with attribute.}
As already mentioned, the ordering algorithms
rely on the fact that random values are uniformly distributed.
However, if the attribute values are not constant but correlated with
the dynamic behavior of the system, the distribution of random values may change
from uniform to  skewed quickly.
For instance, assume that each node maintains an attribute
value that represents its own lifetime. 
Although the algorithm is able to quickly sort random values, so
nodes with small lifetime will obtain the small random values, it
is more likely that these nodes leave the system sooner than other nodes.
This results in
a higher concentration of high random values and a large population of the nodes
wrongly estimate themselves as being part of the higher slices.

\paragraph{Inaccurate slice assignments.}

As discussed in previous sections in detail, slice assignments will
typically be imperfect even when the random values are perfectly
ordered.
Since the ranking approach does not rely on ordering random nodes,
this problem is not raised: the algorithm guarantees eventually
perfect assignment in a static environment.

\paragraph{Concurrency side-effect.}
In the previous ordering algorithms, a non negligible amount of messages are sent unnecessarily.
The concurrency of messages has a drastic effect on the number of useless messages
as shown previously, slowing down convergence.
In the ranking algorithm
concurrency has no impact on convergence speed because all 
received messages are taken in account. 
This is because the information encapsulated in a message (the attribute value of a node)
is guaranteed to be up to date, as long as  the attribute values are constant.

\SubSection{Ranking Algorithm Specification}

The pseudocode of the ranking algorithm is presented in Figure~\ref{alg:attr}.
As opposed to the ordering algorithm of the previous section, the ranking
algorithm does not assign random initial unalterable values as candidate ranks.
Instead, the ranking algorithm improves its rank estimate each
time a new message is received.

The ranking algorithm works as follows.
Periodically each node $i$ updates its view ${\mathcal N}_i$ following 
an underlying protocol that provides a uniform random sample (Line~\ref{N02}); later, we simulate the algorithm using
a variant of Cyclon protocol. See~\cite{FGJKR06b} for further details.
Node $i$ computes its rank estimate (and hence its slice)
by comparing the attribute value of its neighbors to its own attribute value.
This estimate is set to the ratio of the number of nodes with a lower 
attribute value that $i$ has seen over the total number of nodes $i$ has seen (Line~\ref{N14}).
Node $i$ looks at the normalized rank estimate of all its neighbors.
Then, $i$
selects the node $j_{1}$ closest to a slice boundary (according to the rank 
estimates of its neighbors).
Node $i$ selects also a random neighbor $j_{2}$ among its view (Line~\ref{N11}).
When those two nodes are selected, $i$ sends an update message, denoted by a flag $\lit{UPD}$, 
to $j_{1}$ and $j_{2}$
containing its attribute value (Line~\ref{N12}--\ref{N13}).

The reason why a node close to the slice boundary is selected as one of the
contacts is that such nodes need more samples to accurately determine
which slice they belong to (subsection~\ref{sec:analysis2} shows this point).
This technique introduces a bias towards them, so they receive more messages.

Upon reception of a message from node $i$, the passive threads of $j_{1}$ and 
$j_{2}$ are activated so that $j_{1}$ and $j_{2}$ compute their new rank estimate 
$r_{j_{1}}$ and $r_{j_{2}}$.  The estimate of the slice a node belongs to,
follows the computation of the rank estimate.
Messages are not replied, communication is one-way, resulting in identical message
complexity to JK and mod-JK.

\begin{figure}[t]
\centering{
\fbox{
\begin{minipage}[t]{150mm}
\footnotesize
\renewcommand{\baselinestretch}{1.5}
\resetline
\begin{tabbing}
aaaaA\=aaaaaA\=aaaaaaA\kill
{\bf Initial state of node $i$} \\
\line{N00} \> $\ms{period}_{i}$, initially set to a constant;
$r_i$, a value in $(0,1]$; \\
$a_i$, the attribute value;
$b$, the closest slice boundary to node $i$; \\
$g_i$, the counter of encountered attribute values;
$l_i$, the counter \\
of lower attribute values;
$\ms{slice}_i \gets \bot$;
${\mathcal N_i}$, the view.
\\ ~ \\
 
{\bf Active thread at node $i$} \\

\line{N01}  \> $\act{wait}(\ms{period_i})$ \\
\line{N02}  \> $\act{recompute-view}()_i$ \\
\line{N03}  \> $\ms{dist-min} \gets \infty$ \\
\line{N04}  \> {\bf for} $j' \in {\mathcal N}_i$ \\
\line{N05}    \> \T $g_i \gets g_i + 1$ \\
\line{N06}   \> \T {\bf if} $a_{j'} \leq a_i$ {\bf then} $\ell_i \gets \ell_i + 1$ \\
\line{N07} \> \T {\bf if} $\lit{dist}(a_{j'},b) < \ms{dist-min}$ {\bf then} \\
\line{N08} \> \T \T $\ms{dist-min} \gets \lit{dist}(a_{j'},b)$ \\
\line{N09} \> \T \T $j_{1} \gets j'$ \\
\line{N10} \> {\bf end for} \\
\line{N11} \> Let $j_{2}$ be a random node of ${\mathcal N}_i$ \\
\line{N12}  \> $\act{send}(\lit{UPD},a_i)$ to $j_{1}$ \\
\line{N13}  \> $\act{send}(\lit{UPD},a_i)$ to $j_{2}$ \\
\line{N14}   \> $r_i \gets \ell_i / g_i$ \\
\line{N15}   \> $\ms{slice} \gets {\cal S}_{l,u}$ such that $l < r_i \leq u$\\
 ~ \\

{\bf Passive thread at node $i$ activated upon reception} \\
\line{O01} \> $\act{recv}(\lit{UPD},a_j)$ from $j$ \\
\line{O02}   \> {\bf if} $a_{j} \leq a_i$ {\bf then} $\ell_i \gets \ell_i + 1$ \\
\line{O03}   \> $g_i \gets g_i + 1$ \\
\line{O04}   \> $r_i \gets \ell_i / g_i$ \\
\line{O05} \> $\ms{slice} \gets {\cal S}_{l,u}$ such that $l < r_i \leq u$

\end{tabbing}
\normalsize
\end{minipage} 
}
\caption{Dynamic ranking algorithm.}
\label{alg:attr}
}
\end{figure}

\SubSection{Theoretical Analysis}\label{sec:analysis2}

The following Theorem shows a lower bound on the probability 
for a node $i$ to accurately estimate the slice it belongs to.
This probability depends not only on the number of attribute exchanges
but also on the rank estimate of $i$.
For the proof of Theorem~\ref{thm:toprove} please refer to the full version of this paper~\cite{FGJKR06b}.

\begin{theorem}\label{thm:toprove}
Let $p$ be the normalized rank of $i$ and let $\hat{p}$ be its estimate.
For node $i$ to exactly estimate its slice with confidence coefficient of 
$100(1 - \alpha)\%$, the number of messages $i$ must receive is:
$$\left( Z_\frac{\alpha}{2} \frac{ \sqrt{\hat{p}(1-\hat{p})} }{ d } \right)^2,$$
\noindent
where $d$ is the distance between the rank estimate of $i$ and 
the closest slice boundary, and $Z_\frac{\alpha}{2}$ represents the endpoints of the 
confidence interval.  
\end{theorem}

To conclude, under reasonable assumptions all node estimate its slice with confidence coefficient $100(1 - \alpha)\%$, after a finite number of message receipts. 
Moreover a node closer to the slice boundary needs more messages than a node far from the boundary.

\SubSection{Simulation Results}

This section evaluates the ranking algorithm by focusing on three
different aspects.  
First, the performance of the ranking algorithm is
compared to the performance of the ordering algorithm
in a large-scale system where the distribution 
of attribute values does not vary over time.
Second, we investigate if sufficient uniformity is achievable in reality using a dedicated 
protocol.
Third, the ranking algorithm (with and without sliding window technique) and ordering algorithm are compared in 
a dynamic system where the distribution of attribute values may 
change.
For this purpose, we ran two simulations, one for each algorithms.
The system contains (initially) $10^4$ nodes and each view contains $10$ uniformly drawn random 
nodes and is updated in each cycle.
The number of slices is 100, and we present the evolution of the slice disorder measure
over time.

\begin{figure*}[t]
  \begin{center}
    \subfigure[]
    { 
      \label{fig:comparison}    
      \resizebox{2.7in}{!}{\includegraphics[scale=0.5,angle=270,clip=true]{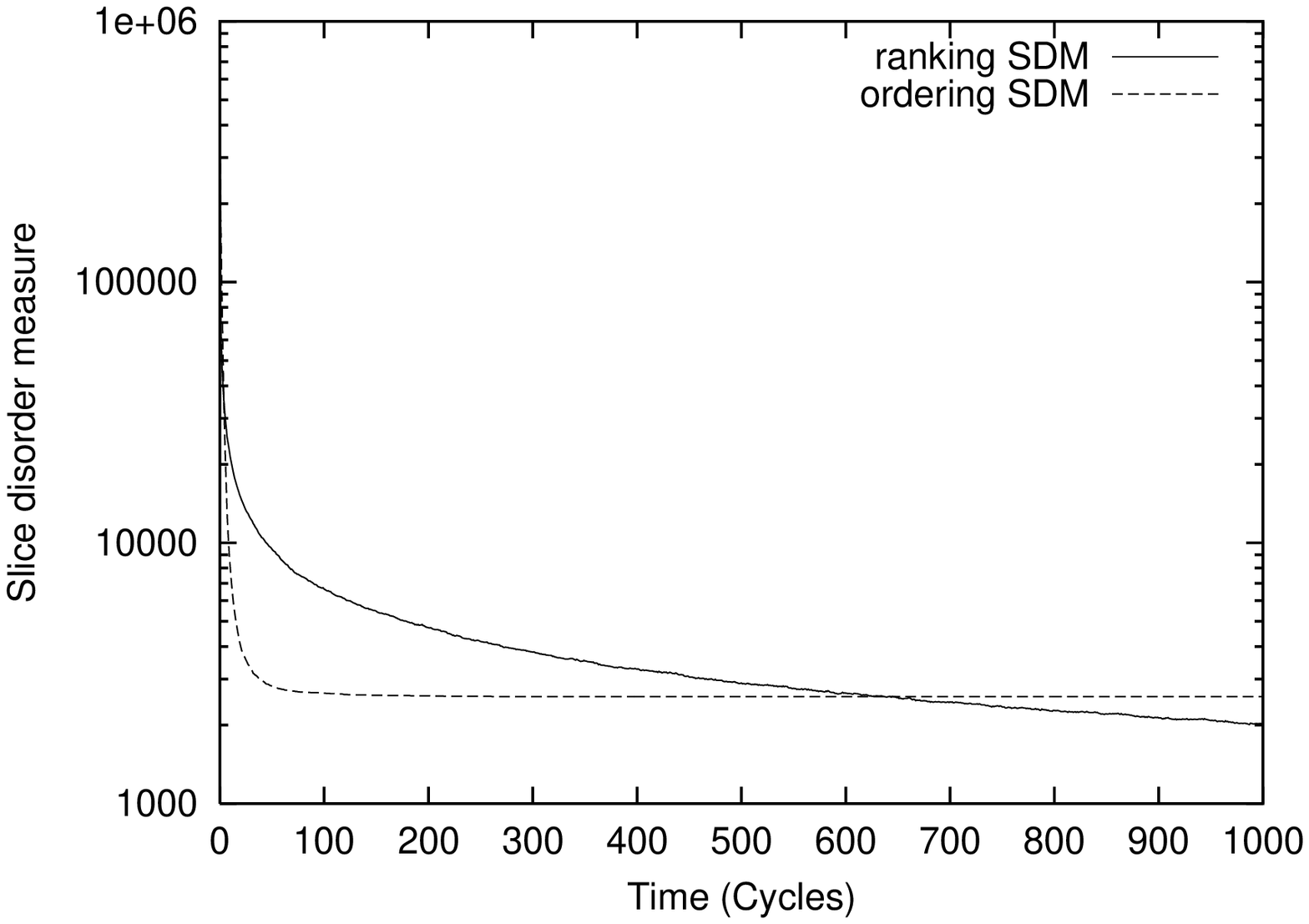}}
    }
    \hspace{1cm}
    \subfigure[]
    { 
      \label{fig:convergence3}
      \resizebox{2.7in}{!}{\includegraphics[scale=0.5,angle=270,clip=true]{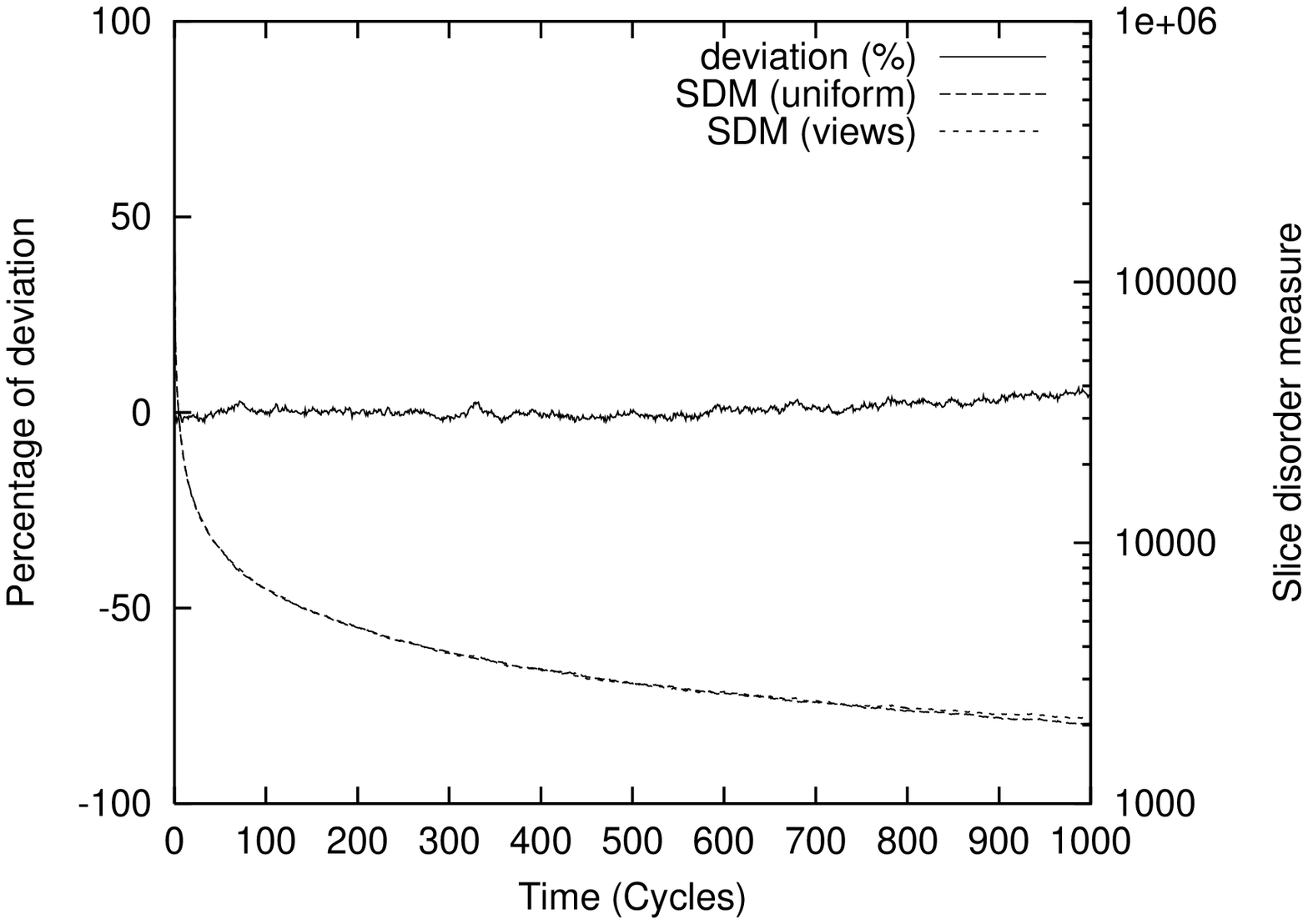}}
    }
    \subfigure[]
    { 
      \label{fig:failures2}
      \resizebox{2.7in}{!}{\includegraphics[scale=0.5,angle=270,clip=true]{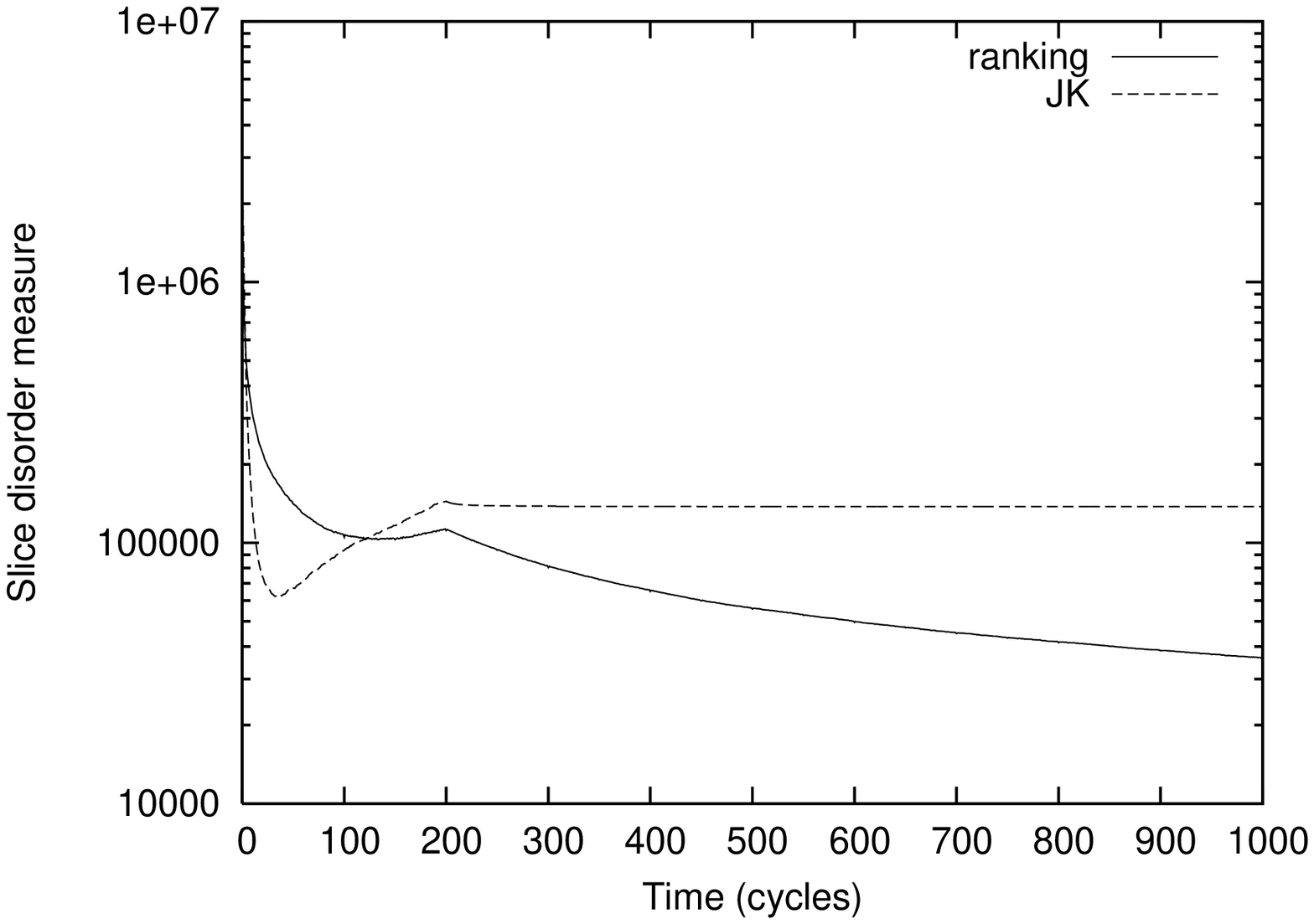}}
    }
    \hspace{1cm}
    \subfigure[]
    {
      \label{fig:failures1}
      \resizebox{2.7in}{!}{\includegraphics[scale=0.5,angle=270,clip=true]{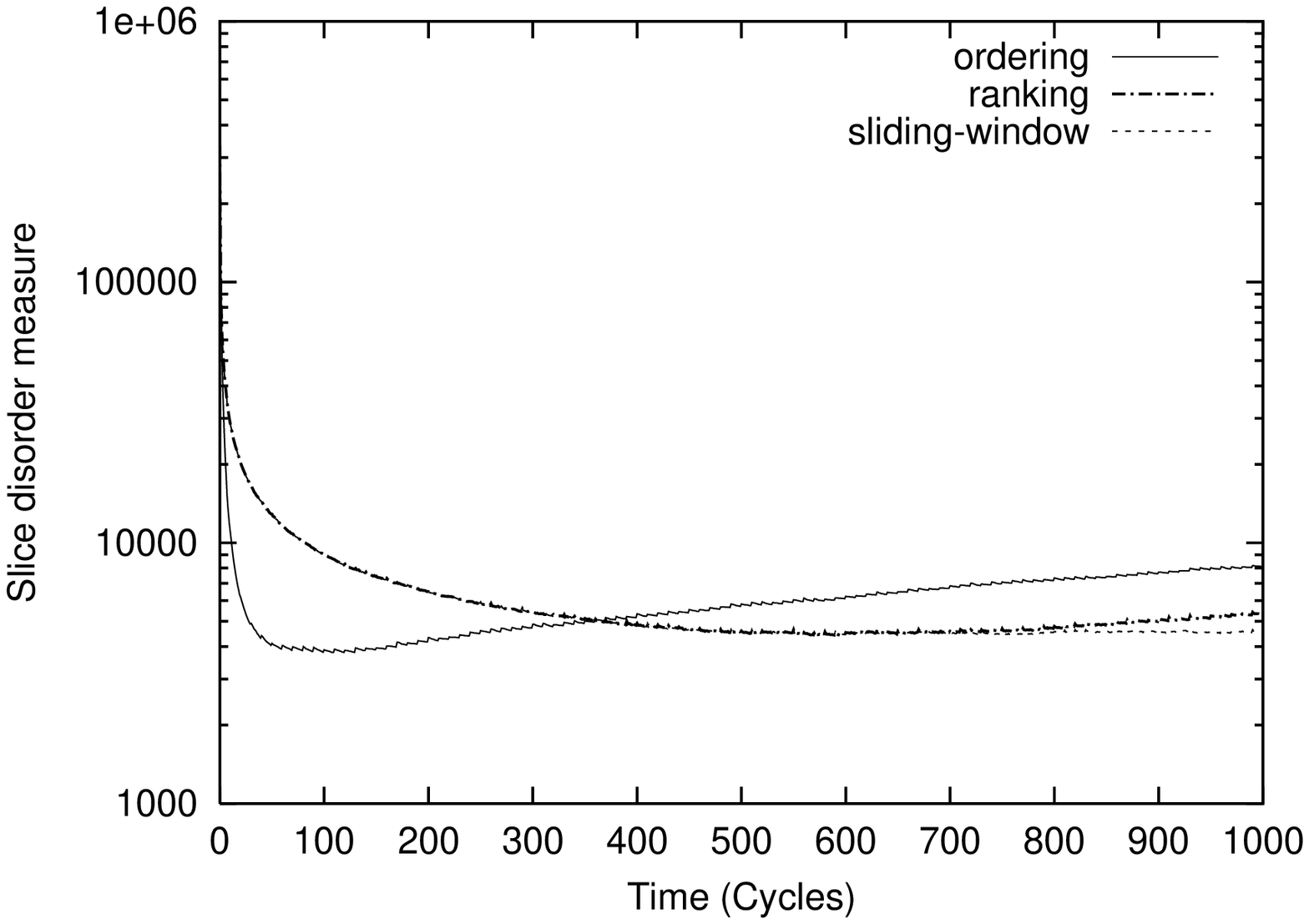}}
    }
    \caption{
       (a) Comparing the ordering algorithm and the ranking algorithms.
       (b) Comparing the uniform drawing and the underlying variant of Cyclon.
       (c) Effect of burst of attribute-correlated churn.
       (d) Effect of a low and regular attribute-correlated churn.
       }
  \end{center}
\end{figure*}


\paragraph{Performance comparison in the static case.}

Figure~\ref{fig:comparison} compares the ranking algorithm to the ordering algorithm
while the distribution of attribute values do not change over time 
(varying distribution is simulated below).
The difference between the ordering algorithm and the ranking algorithm
indicates that the ranking algorithm gives a more precise result (in terms of node to 
slice assignments) than the ordering algorithm.  
More importantly, the slice disorder measure obtained by the ordering algorithm is
lower bounded while the one of the ranking algorithm is not.
Consequently, this simulation shows that the ordering algorithm might fail in 
slicing the system while the ranking algorithm keeps improving its accuracy over 
time.

\paragraph{Feasibility of the ranking algorithm.}

      
Figure~\ref{fig:convergence3} shows that the ranking algorithm does not need artificial
uniform drawing of neighbors.  Indeed, an underlying view management protocol might 
lead to similar performance results.
In the presented simulation we used an artificial protocol, drawing neighbors
randomly at uniform in each cycle of the algorithm execution, and the variant
of the Cyclon view management protocol presented above. 
Those underlying protocols are distinguished on the figure using terms 
"uniform" (for the former one) and "views" (for the latter one).
This figure shows that both cases give very similar results.  
The SDM legend is on the right-handed vertical axis while
the left-handed vertical axis indicates what percentage the SDM difference
represents over the total SDM value.  At any time during the simulation 
(and for both type of algorithms) its 
value remains within plus or minus $7\%$.
The two SDM curves of the ranking algorithm almost overlap.  
To conclude, the use of an underlying distributed protocol that shuffles the view among
nodes can be easily used with the ranking algorithm to provide results similar to the optimal.

\paragraph{Performance comparison in the dynamic case.}

%

In Figure~\ref{fig:failures2} 
the two curves represent the slice disorder measure obtained 
using the ordering algorithm and the ranking algorithm. 
We simulate the churn such that 0.1\% of nodes leave and join 
in each of the 200 first cycles. We observe how the SDM converges. 
The churn is reasonably and pessimistically tuned compared to recent
experimental evaluations~\cite{SR06} of the session duration in three well-known P2P 
systems.

The distribution of the churn is correlated with the attribute values: 
the leaving nodes are the 
nodes with the lowest attribute values while the entering nodes have higher attribute 
values than all nodes already in the system.  
%
The churn introduces a significant disorder in the system which 
counters the fast decrease.  When, the churn stops, the ranking algorithm readapts 
well the slice assignments: the SDM starts decreasing again.  However, 
in the ordering algorithm, the convergence of SDM gets stuck. 
This leads to a poor slice assignment accuracy.

In Figure~\ref{fig:failures1}, 
the three curves represent the slice disorder measure obtained
using the ordering algorithm, the ranking algorithm, and a modified version of the ranking
algorithm with sliding-windows.
(The simulation obtained using sliding windows is described in the next subsection.)
The churn is diminished and made more regular than in the previous simulation such that
0.1\% of nodes leave and join
every 10 cycles.

The decrease slope 
diminishes and the churn effect reduces the amount of nodes with a low attribute 
value while increasing the amount of nodes with a large attribute value.
This unbalance leads to a messy slice assignment, that is, each node must quickly
find its new slice to prevent the SDM from increasing.
%
In the ordering algorithm, the SDM increases faster than in the ranking algorithm.
%
Unlike the ordering algorithm, the 
ranking one keeps re-estimating the rank of each node depending on the 
attribute values present in the system.
Since the churn increases the attribute values present in the 
system, nodes tend to receive more messages with higher attribute values and 
less messages with lower attribute values, which turns out to keep the SDM 
low, despite churn. 
%
To conclude, the results show that when the churn is related to the attribute 
(e.g., attribute represents the session duration, uptime of a node), 
then the ranking algorithm is better suited than the ordering algorithm.

\paragraph{Sliding-window for limiting the SDM increase.}

In Figure~\ref{fig:failures1}, the "sliding-window" curve 
presents 
a slightly modified version of the ranking algorithm
that encompasses SDM increase. 
In 
the ranking algorithm,
upon reception of a new message each node $i$ re-computes
immediately its rank estimate and the slice it thinks it belongs to. 
Consequently the messages
received long-time ago have as much importance as the fresh messages
in the estimate of $i$.
The drawback, as it appeared in Figure~\ref{fig:failures1} of Section~\ref{sec:simu2}, is that if  
the attribute values are correlated with churn, then the precision of the algorithm might diminish.

To cope with this issue, 
upon reception of a message, each node
records an information, about whether the attribute value received 
is larger or lower than the current one. 
Say this information is stored in a first-in first-out buffer such that
only the most recent values remain. 
(One might consider this buffer as a sliding-window.)
Right after having recorded this information, node $i$ can re-compute 
its rank estimate and its slice estimate based 
on the piece of information in the buffer.
Consequently, this improvement increases the algorithm tolerance to change.
%

\Section{Conclusion}
\label{conclusion}

Allocating resources to applications and isolating capable nodes
require specific algorithms that partition the network in a relevant
way. 
The sorting 
algorithm~\cite{JK06} provided a first attempt to 
``slice'' the network. 

In this paper, 
we first proposed the ordering algorithm that improves over this sorting
algorithm.
This improvement comes from  
a judicious choice of candidate nodes to swap values. 
%
Second,
%
we showed that the existing global disorder measure can not
indicate whether nodes find their slice.
That is, we defined the slice disorder measure to measure
how nodes wrongly estimate their slice.
%
Using this new measure, two problems related to the use of 
static random values arose. 
The first one refers to the fact that slice assignment heavily depends on 
the degree of uniformity of the initial random value. 
The second one is related to the fact that
the churn (or failures) might be correlated with the
attribute, leading to 
a wrong slice assignment. 

Last but not least, we provided a ranking algorithm to solve these problems
This algorithm minimizes the effect of correlated churn on slice disorder
and recovers efficiently after a period of correlated churn.
%
%
The convergence speed up of the ordering algorithm and the accuracy of the 
ranking algorithm are proved through theoretical analysis and simulations.

\subsection*{Acknowledgment}
\sloppy{
We wish especially to thank M\'ark Jelasity for the fruitful discussions we had and 
the time he spent improving this paper. 
We are also thankful to Spyros Voulgaris for having kindly shared his 
work on the Cyclon development. 
The work of A.  Fern\'andez and E. Jim\'enez was partially 
supported by the Spanish MEC under grants TIN2005-09198-C02-01, 
TIN2004-07474-C02-02, and TIN2004-07474-C02-01, and the 
Comunidad de Madrid under grant S-0505/TIC/0285. The work of A.  Fern\'andez was done 
while on leave at IRISA, supported by the Spanish MEC under grant 
PR-2006-0193.}

\bibliographystyle{latex8}
\bibliography{slice}

\end{document}